\documentclass[10pt, final, journal, letterpaper, twocolumn]{IEEEtran}

\usepackage{amsfonts}
\usepackage[dvips]{graphicx}
\usepackage{times}
\usepackage{cite}
\usepackage{amsmath}
\usepackage{cases}
\usepackage{array}
\usepackage{dsfont}
\usepackage{amssymb}

\usepackage{stfloats}
\usepackage{slashbox}
\usepackage{graphicx}
\usepackage{footnote}
\usepackage{booktabs}
\usepackage{array}
\usepackage{bm}
\usepackage{color}

\begin{document}

\title{Robust Beamforming for Wireless Information and Power Transmission}

\author{\IEEEauthorblockN{Zhengzheng Xiang and Meixia Tao, \IEEEmembership{Senior~Member,~IEEE}}
\thanks{Manuscript received March 22, 2012; revised May 16, 2012. The associate editor coordinating the
review of this letter and approving it for publication was Dr. Ha Nguyen.}
\thanks{The authors are with the Dept. of Electronic Engineering, Shanghai Jiao Tong University, P. R. China (email:\{7222838, mxtao\}@sjtu.edu.cn).}
\thanks{This work is supported by the Innovation Program of Shanghai Municipal Education Commission under grant 11ZZ19
and the Program for new Century Excellent Talents in University (NCET) under grant NCET-11-0331.}
}
 \maketitle

\begin{abstract}
In this letter, we study the robust beamforming problem for the
multi-antenna wireless broadcasting system with simultaneous
information and power transmission, under the assumption of
imperfect channel state information (CSI) at the transmitter.
Following the worst-case deterministic model, our objective is to
maximize the worst-case harvested energy for the energy receiver
while guaranteeing that the rate for the information receiver is
above a threshold for all possible channel realizations. Such
problem is nonconvex with infinite number of constraints.
Using certain transformation techniques, we convert this problem into
a relaxed semidefinite programming problem (SDP) which can be solved
efficiently. We further show that the solution of the relaxed SDP
problem is always rank-one. This indicates that the relaxation is tight
and we can get the optimal solution for the original problem. Simulation
results are presented to validate the effectiveness of the proposed
algorithm.

\end{abstract}


\section{Introduction}
Energy harvesting for wireless communication is able to extend the
flying power of handheld devices and advocacy for green
communication \cite{Varshney}-\cite{Gupta}. With the aid of this
promising technique, the transmitter can transfer power to terminals
who need to harvest energy to charge their devices, which is
especially important for energy-constrained wireless networks.
Beamforming is another promising technique which exploits channel
state information (CSI) at the transmitter for information
transmission \cite{Cui}-\cite{Wang}. In wireless networks with
simultaneous transmission of power and information, beamforming is
anticipated to play an important role as well.

The beamforming design with perfect knowledge of CSI at the
transmitter was first considered in \cite{Zhangr} to characterize
the rate-energy region in a simplified three-node wireless
broadcasting system. In practical scenarios, perfect knowledge of
CSI may not be available due to many factors such as inaccurate
channel estimation, quantization error, and time delay of the
feedback.

The goal of this letter is to investigate the robust beamformer
design with imperfect CSI for simultaneous information transmission
and energy harvesting. In general, there are two classes of models
to characterize imperfect CSI: the stochastic and deterministic (or
worst-case) models. In the stochastic model, the CSI errors are
often modeled as Gaussian random variables and the system design is
then based on optimizing the average or outage performance
\cite{Verdu}, \cite{Palomar}. Alternatively, the deterministic model
assumes that the CSI uncertainty, though not exactly known, is
bounded by possible values \cite{Luo1}, \cite{WangJH}. In this case,
the system is optimized to achieve a given quality of service (QoS)
for every possible CSI error if the problem is feasible, thereby,
achieving absolute robustness. It was also shown in \cite{Davidson}
that a bounded worst-case model is able to cope with quantization
errors in CSI. In this letter, we shall employ the worst-case
approach to address the robust beamforming design problem.

Consider the three-node system shown in Fig. 1, where we assume that
the transmitter only has imperfect knowledge of the channels to both
the information receiver and energy receiver. We formulate the
worst-case robust beamforming problem for harvested energy
maximization at the energy receiver while ensuring a minimum target
rate at the information receiver. Since the original problem has
infinite constraints due to the channel uncertainties, we first
transform it into an easier problem which has finite constraints but
is still nonconvex. Then we apply the semidefinite relaxation (SDR)
and obtain a semidefinite programming (SDP) problem which can be
solved efficiently. Finally we show that the optimal solution of the
SDP problem is always rank-one, which means that the relaxation is
tight and we can obtain the optimal solution of the original
problem.

The rest of this letter is organized as follows. In Section II, the
system model and the problem formulation are presented. Section III
presents our proposed algorithm to find the solutions to the robust
problems using convex optimization and rank relaxation, and show its
optimality. Simulation results are given in Section IV. Finally,
Section V concludes this letter.

{\bf Notation}: $(\cdot)^\textsl{H}$ and $\mbox{Tr}\{\cdot\}$ stand
for Hermitian transpose and the trace respectively. $|x|$ denotes
the absolute value of the scalar $x$ and $\|\bf{x}\|$ denotes the
Euclidean norm of the vector $\bf{x}$. The function log(.) is taken
to the base $2$.

\begin{figure}
\begin{centering}
\includegraphics[scale=0.6]{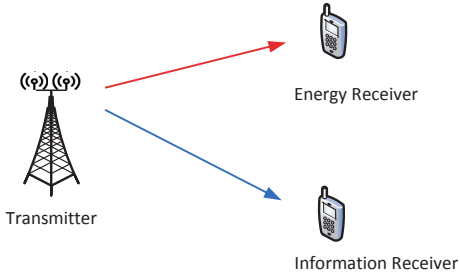}
\vspace{-0.1cm} \caption{A three-node wireless MISO system for
simultaneous information and power transmission.} \label{fig:energy
system}
\end{centering}
\vspace{-0.3cm}
\end{figure}

\section{System Model and Problem Formulation}
With reference to Fig. 1, we consider a three-node multiple-input
single-output (MISO) communication system, where the transmitter has
$N$ antennas and each receiver has a single antenna. Let ${\bf h}^H$
and ${\bf g}^H$ denote the frequency-flat quasi-static $1\times N$
complex channel vectors from the transmitter to the information
receiver and the energy receiver respectively, and $s$ denote the
transmitted symbol. Then the received signals at the information
receiver and the energy receiver are given by, respectively,
\begin{eqnarray}\label{signal model}
&y_i={\bf h}^H{\bf w}s+z_i \\&y_e={\bf g}^H{\bf w}s+z_e
\end{eqnarray}
where ${\bf w}$ is the $N\times 1$ beamforming vector applied to the
transmitter, and $z_i$ and $z_e$ are the additive white circularly
symmetric Gaussian complex noise with variance $\sigma^2/2$ on each
of their real and imaginary components.

For the energy receiver, it will harvest energy from its received
signal. Thanks to the law of energy conservation, we can assume that
the total harvested RF-band power, denoted by $Q$, is proportional
to the power of the received baseband signal, i.e.,
\begin{equation}
Q=\eta|{\bf g}^H{\bf w}|^2
\end{equation}
where $\eta$ is the efficiency ratio at the energy receiver for
converting the harvested energy to electrical energy to be stored.
Here we simply assume that $\eta =1$ and the details for the
converting process is beyond the scope of this letter.

Our objective is to maximize the harvested energy for the energy
receiver while guaranteeing that the information rate for the
information receiver is above a threshold. Mathematically, the
problem is expressed as follows:
\begin{eqnarray}
\label{P0_ob}{\bf P}_0:~ \max\limits_{\bf w}~ |{\bf g}^H{\bf w}|^2~~~~~~~~~~~~~~~~~~~\\
\label{P0_con} ~\mbox{s.t.} ~~\mbox{log}\left(1+\frac{|{\bf
h}^H{\bf w}|^2}{\sigma^2}\right)\geq r\\
 \|{\bf w}\|^2 \leq P~~~~~~~~~~~~~~~~
\end{eqnarray}
where $r$ is the rate target for the information receiver and $P$ is
the power constraint at the transmitter. Similar problem has been considered
in \cite{Zhangr} with the objective of maximizing information rate subject to
a minimum energy threshold.

Herein we consider that the transmitter has imperfect CSI of
both receivers. In particular, the channels are modeled as
\begin{eqnarray}\label{channel}
&{\bf h}=\widehat{{\bf h}}+\Delta{\bf h}\\&{\bf g}=\widehat{{\bf
g}}+\Delta{\bf g}
\end{eqnarray}
where $\widehat{{\bf h}}$ and $\widehat{{\bf g}}$ denote the
estimated CSI known at the transmitter, $\Delta{\bf h}$ and
$\Delta{\bf g}$ are the error vectors. We assume no statistical
knowledge about the error vectors but that they are bounded by some
possible values (also known to the transmitter) as
\begin{eqnarray}\label{error CSI}
&\|\Delta{\bf h}\|\leq \varepsilon \\&\|\Delta{\bf g}\|\leq
\varepsilon
\end{eqnarray}
where $\varepsilon$ is the radius of the uncertainty region.
We assume that both the two receivers have perfect CSI
knowledge.

To take the CSI errors into account, the problem ${\bf P}_0$ based
on worst-case criterion can be formulated as
\begin{eqnarray}
\label{P1_ob}{\bf P}_1:~ \max\limits_{\bf w}\min\limits_{\|\Delta{\bf g}\|\leq \varepsilon} |(\widehat{{\bf g}}+\Delta{\bf g})^H{\bf w}|^2 ~~~~~~~~~~~~~~~~~~~~~~~~~\\
\label{P1_con} ~\mbox{s.t.}
~~\mbox{log}\left(1+\frac{|(\widehat{{\bf
h}}+\Delta{\bf h})^H{\bf w}|^2}{\sigma^2}\right)\geq r, \forall~ \|\Delta{\bf h}\|\leq\varepsilon \\
 \|{\bf w}\|^2 \leq P.~~~~~~~~~~~~~~~~~~~~~~~~~~~~~~~~~~~~~~~~~~
\end{eqnarray}
Since log$(1+x)$ is monotonically increasing for positive $x$,
problem ${\bf P}_1$ can be reformulated as below
\begin{eqnarray}
\label{P11_ob}{\bf P}_1: ~\max\limits_{\bf w}\min\limits_{\|\Delta{\bf g}\|\leq \varepsilon} |(\widehat{{\bf g}}+\Delta{\bf g})^H{\bf w}|^2 ~~~~~~~~~~~~~~~~~~~~~~~~~~\\
\label{P11_con1} \mbox{s.t.}~~ |(\widehat{{\bf h}}+\Delta{\bf
h})^H{\bf w}|^2\geq \sigma^2(2^r-1), \forall \|\Delta{\bf
h}\|\leq\varepsilon~~~
\\ \label{P11_con2}\|{\bf w}\|^2 \leq
P.~~~~~~~~~~~~~~~~~~~~~~~~~~~~~~~~~~~~~~~~~
\end{eqnarray}
It can be seen that the goal of the problem ${\bf P}_1$ is to
maximize the harvested energy for the worst channel realization
while guaranteeing that the information rate is above a threshold
for all possible channel realizations.

\section{Semidefinite Programming Solution}
The key challenges in problem ${\bf P}_1$ are the channel
uncertainties and the nonconvex constraints, which cause that ${\bf
P}_1$ is a semi-infinite nonconvex quadratically constrained
quadratic programming (QCQP) problem. It is well known that the
general nonconvex QCQP problem is NP-hard and thus, intractable.
However, as we will show in the following, due to the special
structure of the objective function and the constraints, problem
${\bf P}_1$ can be reformulated as a convex SDP problem and solved
optimally.

We first transform the above problem into a more tractable form. For
the objective function of ${\bf P}_1$ in \eqref{P11_ob}, we simplify
it using an approach similar to the one developed in \cite{Luo1} and
\cite{Sidiropoulos}. According to triangle inequality, we obtain
\begin{equation}\label{ob_triangle}
|\widehat{{\bf g}}^H{\bf w}+\Delta{\bf g}^H{\bf w}|\geq
|\widehat{{\bf g}}^H{\bf w}|-|\Delta{\bf g}^H{\bf w}|.
\end{equation}
Then applying the Cauchy-Schwarz inequality to the second term in
the right-hand-side (RHS) of \eqref{ob_triangle}, we have
\begin{equation}\label{ob_cauchy}
|\Delta{\bf g}^H{\bf w}|\leq\|\Delta{\bf g}\|\cdot\|{\bf
w}\|\leq\varepsilon\|{\bf w}\|.
\end{equation}
Plugging \eqref{ob_cauchy} into \eqref{ob_triangle}, we then have
that
\begin{equation}\label{triangle_cauchy}
|\widehat{{\bf g}}^H{\bf w}+\Delta{\bf g}^H{\bf w}|\geq
|\widehat{{\bf g}}^H{\bf w}|-|\Delta{\bf g}^H{\bf
w}|\geq|\widehat{{\bf g}}^H{\bf w}|-\varepsilon\|{\bf w}\|.
\end{equation}
An important observation about problem ${\bf P}_1$ is that its
optimal solution is obtained only when the constraint in
\eqref{P11_con2} is active, i.e., the transmitter should work with
full power. Then we have
\begin{equation}\label{ob_objective}
|\widehat{{\bf g}}^H{\bf w}+\Delta{\bf g}^H{\bf w}|\geq
|\widehat{{\bf g}}^H{\bf w}|-\varepsilon\sqrt{P}.
\end{equation}
The inequality becomes equality when $\Delta{\bf g}=-\frac{{\bf
w}}{\|{\bf w}\|}\varepsilon e^{-j\theta}$, where $\theta$ is the
angle between $\widehat{{\bf g}}^H$ and ${\bf w}$. Note that it has
been assumed that $|\widehat{{\bf g}}^H{\bf w}|\geq
\varepsilon\|{\bf w}\|$ in \eqref{triangle_cauchy}, and
$|\widehat{{\bf g}}^H{\bf w}|\geq \varepsilon\sqrt{P}$ in
\eqref{ob_objective}. This assumption essentially means that the
errors $\Delta{\bf g}$ is sufficiently small or equivalently
$\varepsilon$ is sufficiently small. It is a practical assumption
since large channel estimation errors can cause large beamforming
errors and no robustness can be guaranteed in such case. Then
combining \eqref{ob_triangle}-\eqref{ob_objective}, we conclude that
\begin{eqnarray}
\min\limits_{\|\Delta{\bf g}\|\leq \varepsilon} |(\widehat{{\bf
g}}+\Delta{\bf g})^H{\bf w}|^2=\left||\widehat{{\bf g}}^H{\bf
w}|-\varepsilon\sqrt{P}\right|^2.
\end{eqnarray}

For the infinite number of constraints in \eqref{P11_con1}, we can
similarly have that
\begin{equation}
|\widehat{{\bf h}}^H{\bf w}+\Delta{\bf h}^H{\bf w}|\geq
|\widehat{{\bf h}}^H{\bf w}|-\varepsilon\sqrt{P}.
\end{equation}
Here, the equality holds when $\Delta{\bf
h}=-\frac{{\bf w}}{\|{\bf w}\|}\varepsilon e^{-j\varphi}$ with
$\varphi$ being the angle between $\widehat{{\bf h}}^H$ and ${\bf w}$.
Then in order to meet the constraints for all possible $\Delta{\bf
h}$, we just need to satisfy the following
\begin{equation}
|\widehat{{\bf h}}^H{\bf
w}|-\varepsilon\sqrt{P}\geq\sigma\sqrt{2^r-1}.
\end{equation}

Then the robust beamforming problem ${\bf P_1}$ can be rewritten as
follows
\begin{eqnarray}
\label{P111_ob}{\bf P}_1: ~\max\limits_{\bf w} ~|\widehat{{\bf g}}^H{\bf w}|^2 ~~~~~~~~~~~~~~~~~~~~~~~~~~~~~\\
\label{P111_con1} \mbox{s.t.}~~ |\widehat{{\bf h}}^H{\bf w}|^2\geq
\left(\varepsilon\sqrt{P}+\sigma\sqrt{2^r-1}\right)^2
\\ \label{P111_con2}\|{\bf w}\|^2 \leq
P.~~~~~~~~~~~~~~~~~~~~~~~~~~
\end{eqnarray}

Although the problem ${\bf P}_1$ is much easier now, it is still
a nonconvex QCQP problem. We then apply the semidefinite relaxation
and obtain the following relaxed problem:
\begin{eqnarray}
\label{P2_ob}{\bf P}_2: &\max\limits_{\bf W \succeq 0}&\mbox{Tr}\{\widehat{{\bf G}}{\bf W}\}\\
\label{P2_con} &\mbox{s.t.}& \mbox{Tr}\{\widehat{{\bf H}}{\bf
W}\}\geq \left(\varepsilon\sqrt{P}+\sigma\sqrt{2^r-1}\right)^2
\\ & &\mbox{Tr}\{{\bf W}\} \leq P
\end{eqnarray}
where $\widehat{{\bf G}}=\widehat{{\bf g}}\widehat{{\bf g}}^H$ and
$\widehat{{\bf H}}=\widehat{{\bf h}}\widehat{{\bf h}}^H$. Notice
that the rank-one constraint has been dropped and ${\bf P}_2$ is a
relaxed version of ${\bf P}_1$. The problem ${\bf P}_2$ is a
standard SDP problem which is convex and can be solved efficiently
using the software package \cite{Boyd1}.

At this point, an important question is that whether the optimal
solution of ${\bf P}_2$ is rank-one. If ${\bf W}$ is rank-one, then
the optimal beamformer for the original problem ${\bf P}_1$ can be
extracted by eigenvalue decomposition. Otherwise, the
solution of ${\bf P}_2$ is only an upper bound of ${\bf P}_1$ and
the beamformer extracted from ${\bf W}$ is not guaranteed to be
globally optimal. Generally there is no guarantee that an algorithm
for solving SDP problems will give the desired rank-one solution.
However, in some special cases such as \cite{Ottersten}-\cite{Huang}, the relaxation is
proven to be exact and thus there always exists a rank-one solution.
Whether the relaxation is tight for our proposed algorithm will be
addressed in the following theorem.

{\bf Theorem 1:} The optimal solution $\bf W$ for problem ${\bf
P}_2$ is rank-one.
\begin{proof}
Please refer to Appendix A.
\end{proof}

According to Theorem 1, we can see that problem ${\bf P}_2$ is
indeed equivalent to the original problem ${\bf P}_1$, which means
that the relaxation is tight. So in order to solve the problem ${\bf
P}_1$, we first solve the SDP problem ${\bf P}_2$ and obtain the
resulting rank-one matrix ${\bf W}^\star$. Apply eigenvalue decomposition on ${\bf
W}^\star$ as
\begin{equation}\label{EVD_W}
{\bf W}^\star=\alpha^\star{\bf w}^\star{{\bf w}^\star}^H.
\end{equation}
The optimal solution of ${\bf P}_1$ is then obtained as ${\bf
w}=\sqrt{\alpha^\star}{\bf w}^\star$.

\begin{figure}
\begin{centering}
\includegraphics[scale=0.4]{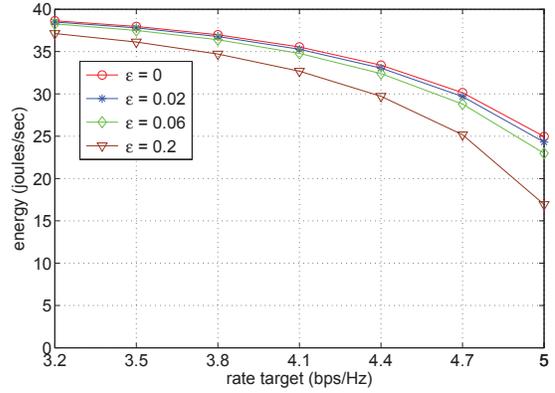}
\vspace{-0.1cm} \caption{Average harvested energy for the robust
beamforming design.} \label{fig:robust}
\end{centering}
\vspace{-0.3cm}
\end{figure}

\begin{figure}
\begin{centering}
\includegraphics[scale=0.4]{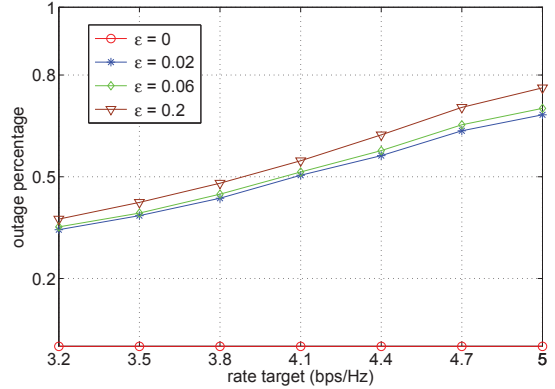}
\vspace{-0.1cm} \caption{Outage percentage for the nonrobust
beamforming design.} \label{fig:nonrobust}
\end{centering}
\vspace{-0.3cm}
\end{figure}

\section{Simulation Results}
In this section, we present numerical results to evaluate the
performance of the proposed robust beamforming algorithm. We consider
the three-node MISO system in which the transmitter has
four antennas $(N = 4)$. We set the power $P=10$ and noise covariance
$\sigma^2=1$. The channel from the transmitter to each receiver
is assumed as the normalized Rayleigh fading channel. The rate target
for the information receiver is set to  be smaller than
$\log(1+10\|{\bf h}\|^2)$ in order for the problem to be feasible.
For simplicity, we normalize these channel vectors with respect to the number of transmit antennas as $\|{\bf h}\|^2=\|{\bf g}\|^2=4$. Thus the
feasible region for the rate target is $0\leq r\leq \log(1+40)$. A
total of $100$ independent normalized channel realizations are simulated. For
each channel realization, $100$ channel uncertainty samples are
generated.

In Fig. 2, we plot the average harvested energy versus
different targets of information rate for different levels of
bounded channel uncertainty. The special case with the CSI (i.e.,
${\varepsilon=0}$) is also simulated. It can be seen that the performance
loss is small when the CSI error is not big. Also the performance gap increases
when the rate target becomes larger.

In order to show how important it is to take the channel uncertainty
into account when designing the beamformers, let us assume that the
beamforming design takes place under the assumption of perfect CSI
at the transmitter while in fact there is some uncertainty
associated with the CSI used in the design problem, which we call
the ``nonrobust beamforming design". Fig. 3 shows percentage of outage
\footnote{We call the outage happens when the rate target is
not satisfied at the information receiver.} at different rate targets for the nonrobust design. We observe that the
channel uncertainty, when not considered in the design process,
leads to frequent violations of the rate target at the information
receiver. However, for our proposed worst-case robust beamforming
algorithm, the rate target is always satisfied and no outage
happens.

\section{Conclusion}
In this letter, we consider the worst-case robust beamforming design
for the wireless communication system with both information and
energy receivers when the CSI is imperfect. By means of semidefinite
relaxation, we transform the original robust design problem into a
SDP problem. Then we prove that such
relaxation is tight and we can always obtain the optimal solution of
the original problem. The performance of the proposed beamforming
algorithm has been demonstrated by simulations. Future research
directions may include the robust beamforming design for the more
general broadcasting systems with multiple information receivers and
multiple energy receivers.

\appendices
\section{Proof of Theorem 1}
Denote
$\beta\triangleq\left(\varepsilon\sqrt{P}+\sigma\sqrt{2^r-1}\right)^2
$, the Lagrangian of ${\bf P}_2$ is given by
\begin{eqnarray}
L\left({\bf W},\lambda,\mu\right)=\mbox{Tr}\{\widehat{{\bf G}}{\bf
W}\}+\lambda\left(\mbox{Tr}\{\widehat{{\bf H}}{\bf W}\}-
\beta\right)\\\nonumber-\mu\left(\mbox{Tr}\{{\bf
W}\}-P\right),~~~~~~~~~~~~~~
\end{eqnarray}
where $\lambda$ and $\mu$ are the dual variables. The Lagrange dual
function is then defined as
\begin{equation}
g(\lambda,\mu)=\max\limits_{\bf W\succeq 0}L\left({\bf
W},\lambda,\mu\right).
\end{equation}
Since ${\bf P}_2$ is convex with strong duality, we can solve it by
solving its dual problem
\begin{equation}
{\bf D}_2:\min\limits_{\lambda\geq0,\mu\geq0}g(\lambda,\mu).
\end{equation}

Denote the optimal solution of ${\bf D}_2$ as
$(\lambda^\star,\mu^\star)$, then the matrix ${\bf W}^\star$ that
maximizes $L\left({\bf W},\lambda^\star,\mu^\star\right)$ is the
optimal solution of ${\bf P}_2$, which means that we can find ${\bf
W}^\star$ through the following problem
\begin{equation}\label{find_S}
\max\limits_{\bf W\succeq 0}~\mbox{Tr}\{\widehat{{\bf G}}{\bf
W}\}-\mbox{Tr}\left\{\left(\mu^\star{\bf
I}-\lambda^\star\widehat{{\bf H}}\right){\bf W}\right\}
\end{equation}
where the constant terms has been discarded. In order for problem
\eqref{find_S} to have a bounded value, it is shown as follows that
the matrix $\mu^\star{\bf I}-\lambda^\star\widehat{{\bf H}}$ should
be positive definite. Suppose $\mu^\star{\bf
I}-\lambda^\star\widehat{{\bf H}}$ is not positive definite, then we
can choose ${\bf W}=t{\bf w}{\bf w}^H$, where $t>0$ and
$\mbox{Tr}\{t(\mu^\star{\bf I}-\lambda^\star\widehat{{\bf H}}){\bf
w}{\bf w}^H\}\leq 0$. Due to the independence of $\widehat{{\bf g}}$
and $\widehat{{\bf h}}$, it follows that $\mbox{Tr}\{t\widehat{{\bf
G}}{\bf w}{\bf w}^H\}>0$ with probability one. Let $t\rightarrow
+\infty$, the optimal value in \eqref{find_S} will be unbounded,
which is a contradiction of the optimality of
$(\lambda^\star,\mu^\star)$.

Define ${\bf Q}\triangleq( \mu^\star{\bf
I}-\lambda^\star\widehat{{\bf H}})\succ{\bf 0}$ and let
$\overline{{\bf W}}={\bf Q}^{1/2}{\bf W}{\bf Q}^{1/2}$, the problem
in \eqref{find_S} is then rewritten as
\begin{equation}\label{find_S_1}
\max\limits_{\overline{{\bf W}}\succeq 0}~({\bf
Q}^{-1/2}\widehat{{\bf g}})^H\overline{{\bf W}}({\bf
Q}^{-1/2}\widehat{{\bf g}})-\mbox{Tr}\{\overline{{\bf W}}\}.
\end{equation}
Then we claim that the optimal solution of \eqref{find_S_1} is
always rank-one. Suppose the optimal solution ${\overline{{\bf
W}}^\star}$ is not rank-one, without loss of generality, we can
assume its rank is $k$ ($2\leq k\leq N$) and decompose it as
${\overline{{\bf W}}^\star}=\sum_{j=1}^k\alpha_j\overline{{\bf
w}}_j\overline{{\bf w}}_j^H$. Then we choose another
${\overline{{\bf W}}^\star}'=(\sum_{j=1}^k\alpha_j)\overline{{\bf
w}}_i\overline{{\bf w}}_i^H$, where
$i=\mbox{arg}\max\limits_{j\in\{1,...,k\}}|({\bf
Q}^{-1/2}\widehat{{\bf g}})^H\overline{{\bf w}}_j|$. Then
${\overline{{\bf W}}^\star}'$ can achieve a larger value than
${\overline{{\bf W}}^\star}$, which is a contradiction.

From the above discussions, it is known that ${\overline{{\bf
W}}^\star}$ is always rank-one. Since ${\bf W}^\star={\bf
Q}^{-1/2}{\overline{{\bf W}}^\star}{\bf Q}^{-1/2}$, we must have
that ${\bf W}^\star$ is rank-one, which completes the proof of
Theorem 1.

\end{document}